# A Critical Review of Concepts, Benefits, and Pitfalls of Blockchain Technology Using Concept Map

**IYOLITA ISLAM[1], KAZI MD. MUNIM[1], SHAHRIMA JANNAT OISHWEE[1],
A. K. M. NAJMUL ISLAM[2], AND MUHAMMAD NAZRUL ISLAM[1]**
[1]Department of Computer Science and Engineering, Military Institute of Science and Technology (MIST), Mirpur Cantonment, Dhaka-1216, Bangladesh
[2]Department of Future Technologies, University of Turku, 20500 Turku, Finland

Corresponding author: A. K. M. Najmul Islam (najmul.islam@utu.fi)

**ABSTRACT** Blockchain is relatively a new area of research. However, a surge of research studies on the blockchain has taken place in recent years. These research studies have mostly focused on designing and developing conceptual frameworks to build more reliable, transparent and efficient digital systems. While blockchain brings a wide variety of benefits, it also imposes certain challenges. Therefore, the objective of this research is to understand the properties of blockchain, its current uses, observed benefits and pitfalls to provide a balanced understanding of blockchain. A systematic literature review approach was adopted in this paper in order to attain the objective. A total of 51 articles were selected and reviewed. As outcomes, this research provides a summary of the state-of-the-art research studies conducted in the area of blockchain. Furthermore, we develop a set of concept maps aiming to provide in-depth knowledge on blockchain technology for its efficient and effective usage in the development of future technological solutions.

## I. INTRODUCTION

Blockchain technology is one of the latest innovations, which can be considered as a new paradigm for the regulation of human and business activities [1]. It is a distributed consensus mechanism to store the transaction information in a Peer-to-Peer (P2P) network. It was actually designed as an open-source project to introduce a digital currency (i.e. cryptocurrency) named bitcoin [1]. Although the concept of blockchain was first discussed through bitcoin, it has use cases that go far beyond the cryptocurrencies.

Blockchains can be either public or private [2]. Public blockchains are permissionless and therefore, anyone can join. Bitcoin is a public blockchain. In contrast, in order to join a private blockchain, permission is needed. IBM's Hyperledger is an example of private blockchain. The first generation blockchain (i.e. Bitcoin) did not support smart contracts and was capable to store data related to only transactions of bitcoin. A smart contract is a piece of code that

The associate editor coordinating the review of this manuscript and approving it for publication was Aneel Rahim.

execute predefined actions when certain conditions within the system are met. Blockchain platforms such as Ethereum and IBM Hyperledger support smart contracts. These platforms are also able to store any type of data. Therefore, recently blockchain is used in many domains (other than cryptocurrencies) to achieve more security, flexibility, efficiency and transparency. The domains include health informatics for storing and managing patient data [3], [4], secured energy trading [5], banking and financial sector to facilitate transparent transactions [6]–[8], e-governance to improve government services [9], development of smart city [10], [11], [12], internet of things (IoT) services integrated with blockchain [13], [14], decentralized and auditable software validation system [15], among others. In other studies [1], [16], [17], some possible threats of implementing blockchain technology are highlighted. These include for example, wasted resources such as electricity and storage, scalability, lack of usability, and privacy, among others. Therefore, it becomes very crucial to understand the in-depth concept of blockchain and its applications (or context of use) with respect to the possible benefits and threats.

The objective of this research is to provide the fundamental concept of blockchain for enhancing its effective usage in the development of future technological solutions. To attain this objective, this research presents a set of concept maps [18] to provide an in-depth definition of blockchain, all possible features of blockchain and the association among the properties, benefits, and pitfalls of blockchain.

We use the concept map as it allows us to structure, organize, and represent knowledge [18]. In other words, a concept map can construct ideas into a hierarchical structure of concepts [19]. Understanding the 'concept' and 'proposition' are the primary requirements to construct a good concept map or to learn about the concept map. The term 'concept' refers to the perceived regularity in events/objects while the 'proposition' refers to the statements about an object/event in the universe that may occur naturally or be constructed. Moreover, the concept maps are used to facilitate the study/learning procedure, research, evaluation and data analysis [20], [21]. Therefore, the concept maps facilitate to provide fundamental knowledge on a specific subject.

The rest of the article is organized as follows. Section II presents related works, section III presents the method of the study, section IV explains the blockchain technology and its features, benefits and pitfalls, section V provides a mapping among the features, advantages and pitfalls of blockchain and finally section VI presents the final outcomes and limitations of this study with future expansion possibilities.

## II. LITERATURE REVIEW

This section briefly discusses the related work that has been conducted on the working principles, applications, benefits, and limitations of blockchain.

Davidson [6] characterized blockchain as a catallaxy for being a robust, protected and transparent ledger since it implements secured mechanism using cryptography. According to Crosby [22], blockchain is a distributed online database of all digital events occurred among the participant nodes in a network. He provided an overview of blockchain technology and described some challenges, which can be overcome by blockchain and some limitations to be resolved in future work. Buterin [23] referred blockchain as a crypto-economically secured magic computer that includes self-executable programs with records of all previous and current states. Carlozo [24] described blockchain technology as the backbone of each digital transaction. He also asserted that blockchain would offer more dynamic approaches to business.

A systematic review conducted by Yli [17] highlighted the recent research developments using blockchain and provided the possible future research directions that include introducing a new cryptocurrency, usage of multi-level authentication techniques, and energy-efficient resource management for distributed systems, among others.

Ammous [25] analyzed the working principle of blockchain technology in several domains to show how blockchain provides benefits such as flexibility, tamper-resistant and automated validation in a system. Beck [26] stated that blockchain is more appropriate to facilitate complex business transactions and developing new business models.

**TABLE 1.** Proposition table for concept map to define blockchain.

| Ser | Concept | Relation | Concept | Ref |
|---|---|---|---|---|
| 1 | Blockchain | *consists of* | Blocks | [7] [15] [22] [24] [28] |
| 2 | Blocks | *contain* | Messages | [15] [24] [28] |
| 3 | Blocks | *contain* | Proof of Work | [15] |
| 4 | Blocks | *contain* | Reference | [15] [24] [28] |
| 5 | Blockchain | *maintains* | Historical Record | [7] [15] [22] |
| 6 | Historical Record | *preserves* | Irreversibility of Record | [7] |
| 7 | Blockchain | *maintains* | Transparency | [7] |
| 8 | Transparency | *uses* | Pseudonimity | [7] |
| 9 | Blockchain | *includes* | Shared Database | [7] [22] [24] [26] |
| 10 | Shared Database | *secures* | Cryptocurrecy | [4] [26] |
| 11 | Blockchain | *includes* | Transaction | [7] |
| 12 | Transaction | *made in* | Cryptocurrecy | [4] [7] [26] |
| 13 | Transaction | *made in* | Bitcoin | [7] [10] [11] |
| 14 | Blockchain | *includes* | Peer to Peer Network | [7] |
| 15 | Peer to Peer Network | *used in* | Bitcoin | [7] |

Some researchers worked to find the threats of adopting blockchain technology. Swan [1] represented some technical challenges in terms of throughput, latency, security, and usability, among others. Amosova [27] analyzed the risks of unregulated use of blockchain technologies in the financial market from the perspectives of law enforcement agencies, financial institutions, civil societies, individual persons, and regulators.

Pilkington [28], [29] briefly discussed the categories, features and working principles of blockchain technology. He depicted some areas of application, for example, gridcoin (a system for grid computing), providing digital ID, voting system, and banking using the smart contracts. Bohme [30] described the centralization and decentralization concept of bitcoin, its usage, merits, and demerits.

In sum, the earlier studies emphasized to present the technological progress, applications, and threats of blockchain in different fields. No review study was found that explicitly focused to present all possible features of blockchain. Similarly, no study was found that depicts a clear mapping among the identified benefits, pitfalls and context of use. Thus, this research focus to provide the fundamental concept of blockchain technology through concept maps for the optimal usage of blockchain in future technological progress.

## III. RESEARCH METHOD

In this research a systematic literature review procedure [31] was adopted to attain the research aim. For selecting the primary articles, the major databases such as IEEE Explorer, Springer Link, ACM digital library, Science Direct, and Google Scholar were searched for related articles. The search keywords used for finding materials were: ''Blockchain'', ''Blockchain Technology'', ''Applications of Blockchain'', ''Concept of Blockchain'', ''Benefits of Blockchain'', ''Pitfalls of Blockchain'', ''Blockchain and concepts'', ''Blockchain and properties'', ''Blockchain and advantages'', and ''Blockchain and pitfalls''. These strings were applied for all the above-mentioned scholar databases as well as Google search engine. A total of 860 articles were found initially. Several screenings were performed to select the best-matched papers. The criteria: the research materials were from the year 2014 or afterwards; the papers were published in conference proceedings, magazines, journals or books; and the papers were written in English, were used. We also excluded duplicate articles. After that, by reviewing the title, abstract and introduction, 51 articles were selected for review.

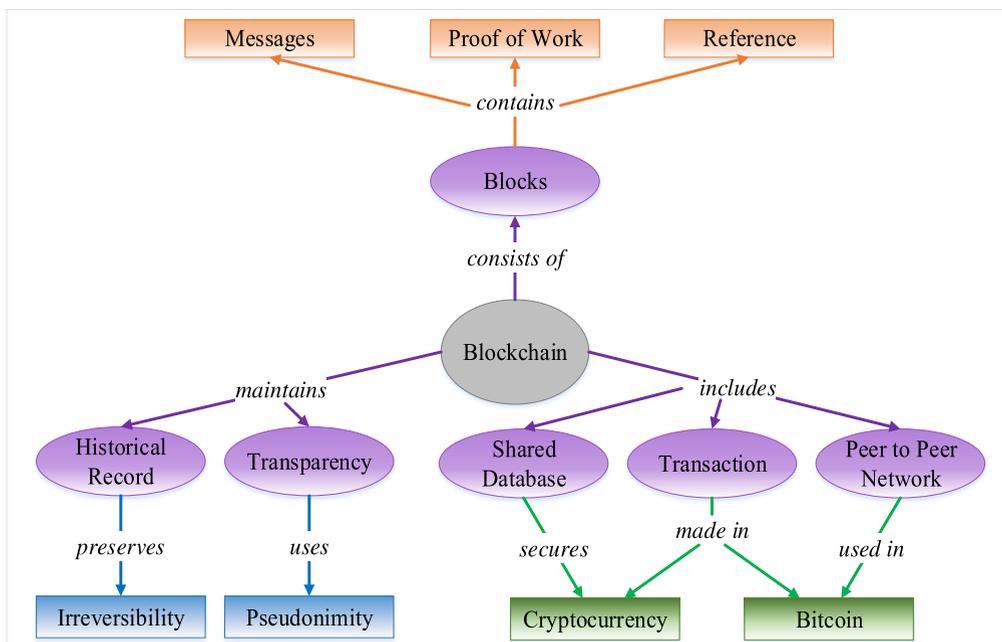

**FIGURE 1.** Concept map of blockchain.

**TABLE 2.** Properties of blockchain.

| Ser | Properties | Description | Ref | Freq |
|---|---|---|---|---|
| 1 | Shared Database | Blockchain is implemented using shared databases, which are logically linked among a distributed network. Participants in a distributed network collaborate to agree on the true state of the database. There is no central or single party as controller. | [4] [7] [13] [22] [28] [32] [33] [34] [35] [36] [37] | 11 |
| 2 | P2P Transmission | Each of the participants in a network are called node. User can transfer or share data between nodes in duplex mode. | [7] [33] [35] [36] [37] [38] [39] [40] | 8 |
| 3 | Timestamped Blocks | The blocks are linked sequentially by the time when they are generated. | [4] [24] [34] [35] [41] [42] [43] [44] | 8 |
| 4 | Immutable Records | Once any block has been created, they are irreversible and non-updatable. Because each block has the hashed value of the information from the previous block and a time-stamp. | [4] [7] [24] [29] [35] [44] | 6 |
| 5 | Encrypted Data Transmission | Data is encrypted by sender's public key and decrypted by private key of the receiver for secured transmission. | [4] [28] [30] | 3 |
| 6 | Disintermedia-tion | Blockchain is executed in a distributed network, which uses a proof-of-work consensus mechanism. It doesn't require any third party. So, the dependency on any third party has been removed. | [3] [16] [33] [35] [44] [45] [46] | 7 |
| 7 | Computational Logic | Blockchain can include computational logic implementation. User can introduce rules and algorithms (i.e. smart contracts), which can be automatically executed while data transmission. | [7] [36] [37] [47] | 4 |
| 8 | Transaction Dependency | Transactions are performed depending on some pre-specified rules and regulation. | [36] [37] [48] [49] | 4 |
| 9 | Transaction Rules | There are some transaction rules declared by smart contracts. | [50] [51] [52] | 3 |
| 10 | Distributed Trust | Each user has a unique address. Every transaction information is visible to all the participants over the network. | [7] [36] [37] [44] | 4 |
| 11 | Multiple Writers | From Blockchain multiple companies can search reliable data. These data are supplied by multiple writers in the shared databases. | [53] [54] [55] [56] | 4 |
| 12 | Validation | When any block is created, it needs to be verified by other participants over the network. This is called mining in blockchain. | [3] [4] [22] [44] | 4 |
| 13 | Scalability | Blockchain is an expansible database. A blockchain can expand without any limit. | [32] [33] [34] [35] [57] [58] | 6 |

**TABLE 3.** Proposition table for concept map to represent the properties of blockchain.

| Ser | Concept | Relation | Concept | Relation | Concept |
|---|---|---|---|---|---|
| 1 | | | Service Perspective | *which offers* | Scalability |
| 2 | | *includes* | | *which has* | Validation |
| 3 | | | | *which offers* | Multiple Writers |
| 4 | | | | *which has* | Distributed Trust |
| 5 | | *supports* | Logical Inclusion | *which provides* | Computational Logic |
| 6 | Blockchain | | | *which has* | Transaction Dependency |
| 7 | | | | *which offers* | Transaction Rules |
| 8 | | | Architectural Characteristics | *which have* | Shared Database |
| 9 | | | | *which use* | P2P Transmission |
| 10 | | *includes* | | *which comprehend* | Disintermediation |
| 11 | | | | *which have* | Timestamped Blocks |
| 12 | | | | *which include* | Immutable Records |
| 13 | | | | *which use* | Encrypted Data Transmission |

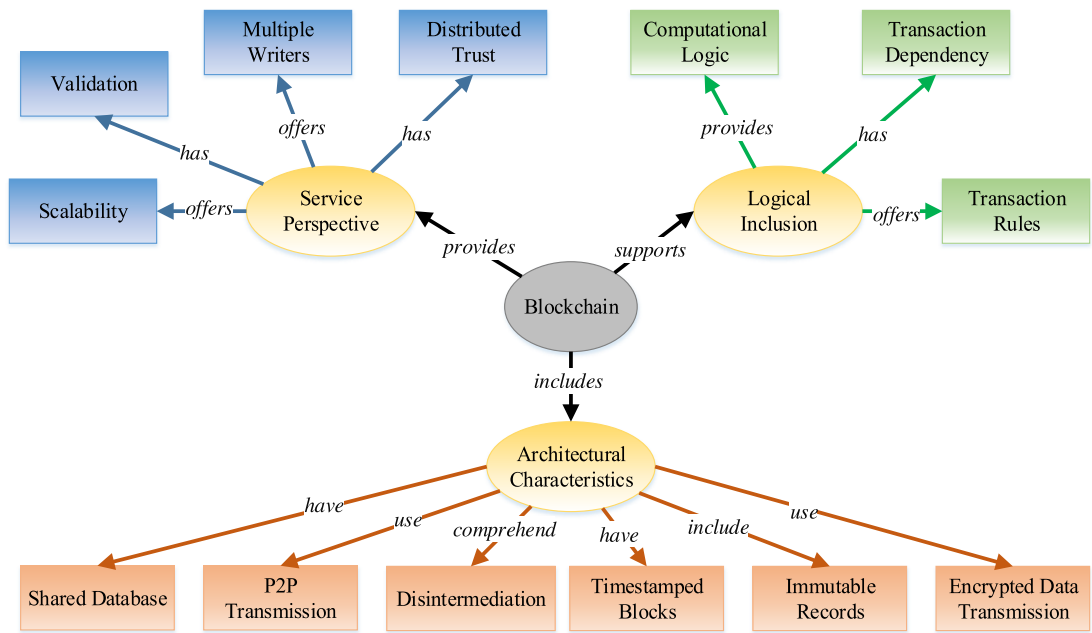

**FIGURE 2.** Conceptual map of features of blockchain.

While reviewing each of the article, a set of data were extracted that includes: the aim of the research, application domain or context of use, features or properties, advantages, and pitfalls of blockchain. The extracted data were then analyzed and synthesized to present a set of concept maps for providing an in-depth understanding of blockchain.

## IV. BLOCKCHAIN TECHNOLOGY

Blockchain is a chain of blocks where the blocks are linked using cryptography. Each block contains the hash of the previous block, a time-stamp, proof-of-work signature, and transaction data.

To add a new block to the blockchain, the following events are required to happen. Firstly, a transaction must occur in the network. Secondly, the details of the transactions must be verified by the participants (i.e. miners) over the network. Thirdly, after a transaction is verified by the miners, data are stored in the block. Finally, the block is linked by a hashed value of the information from the previous block. The new block is added to the blockchain and becomes public to others.

## A. DEFINITION OF BLOCKCHAIN

The literature survey showed that blockchain consists of blocks that contain messages, proof of work and reference of the previous block. Furthermore, shared databases, transactions and P2P network are associated with blockchain. Shared databases are secured while the transactions mainly refer to the transfer of a cryptocurrency like bitcoin in a P2P network. Blockchain maintains the historical record to maintain transparency. Historical records are irreversible. Transparency is ensured by using pseudonymous transactions. These propositions define the blockchain and are presented in Table 1. A concept map grounded on the propositional table is depicted in Fig. 1. Referring to the concept map, a formal definition of blockchain could be derived as:

> *Blockchain consists of blocks containing messages, proof of work and reference of the previous block and stored in shared database, which*

> *is able to perform transactions over P2P network maintaining irreversible historical records and transparency.*

**TABLE 4.** Advantages of blockchain.

| Ser | Advantage | Description | Context | Ref | Freq |
|---|---|---|---|---|---|
| 1 | Independence from third party | Blockchain uses a powerful mechanism with cryptoeconomy to authenticate each transaction. This makes the system trustless. So, it doesn't require any third party. | IoT (Smart City) & Economics (Bitcoin) & Medical | [3] [6] [11] [22] | 4 |
| 2 | Enhanced Security | Transaction uses cryptography. So, all the data transfers are secured. | Finance & IoT (Smart City) & SLV | [7] [11] [15] | 3 |
| 3 | Flexible & efficient | Easy to modify and integrate in existing system. | Economics | [6] [7] [40] | 3 |
| 4 | Auditable & verifiable | Keeping track of every transaction makes the system transparent to all users. The records are auditable and verifiable. | Economics (Bitcoin) & SLV & Medical | [3] [15] [22] | 3 |
| 5 | Robust | System can recover from attacks or any failure. Man-in-the middle attack is quite impossible in this transparent system. | Economics & Finance | [6] [7] [40] | 3 |
| 6 | Reduced Cost | Blockchain implementation eliminates the cost for commission or any other function. So, it offers low transaction cost. | Finance & Medical | [3] [7] | 2 |
| 7 | Auto Synchronized | It is not required to receive the transaction in the same serial in which they were generated. | Economics (Bitcoin) & SLV | [15] [22] | 2 |
| 8 | Less Redundant | Reduced redundancy due to peer-to-peer collaboration | Finance | [7] | 1 |
| 9 | Reliable | Achieving trust of the users | Medical | [4] | 1 |
| 10 | Multi-accessible | Multiple users can access to a system at the same time | Finance | [7] | 1 |

**TABLE 5.** Pitfalls of blockchain.

| Ser | Pitfalls | Description | Context | Ref | Freq |
|---|---|---|---|---|---|
| 1 | Complicated Usability | Implementing the bitcoin API or general blockchain application platform for developing services is quite complex. | E-government | [1] [13] [17] | 3 |
| 2 | Legalization | There is a lack of legislation and regulation. | E-government & Economics | [6] [13] [26] | 3 |
| 3 | Wasted Resources | Bitcoin wastes a humongous amount of energy for mining. The energy loss is approximately $15million/day. | Economics | [1] [17] [58] | 3 |
| 4 | Latency | Bitcoin blockchain takes minimum 10 minutes to ensure sufficient security for each transaction. | Economics | [1] [17] [33] | 3 |
| 5 | Less Throughput | Throughput in bitcoin blockchain is 1.00-7.00 tps (transaction per second). VISA has 2000-5000 tps and Twitter has 5000-15000 tps | Economics | [1] [17] | 2 |
| 6 | Contractual Enforcement | If any voluntary contract is sanctioned by the central government and not connected with the shadow of law, the problem of contractual enforcement arises. | Economics | [1] [6] | 2 |
| 7 | Inadequate Security | Chance of Cyber attack in Blockchain is very high. | Economics | [1] [17] | 2 |
| 8 | Non Erasable | Once written, data cannot be deleted from Blockchain | Economics (Bitcoin) | [22] | 1 |
| 9 | Anonymous Transactions | Without any governmental control, Bitcoin can make a multibillion dollar transaction, which can create a great controversy. | Economics (Bitcoin) | [22] | 1 |
| 10 | Identification Crisis | The addresses have no name or other customer information, which creates identity crisis in bitcoin transaction. | Finance | [26] | 1 |

### B. PROPERTIES OF BLOCKCHAIN

A total of thirteen properties or features were retrieved through our literature survey. The properties are briefly represented in Table 2. The mostly (eleven times) stated property was *shared database* while *timestamped blocks* and *P2P transmission* were articulated in eight articles.

Later, based on the thematic and logical relations, the retrieved properties were categorized into three clusters through affinity diagram [59]. The clusters are service perspective, logical inclusion, and architectural characteristics.

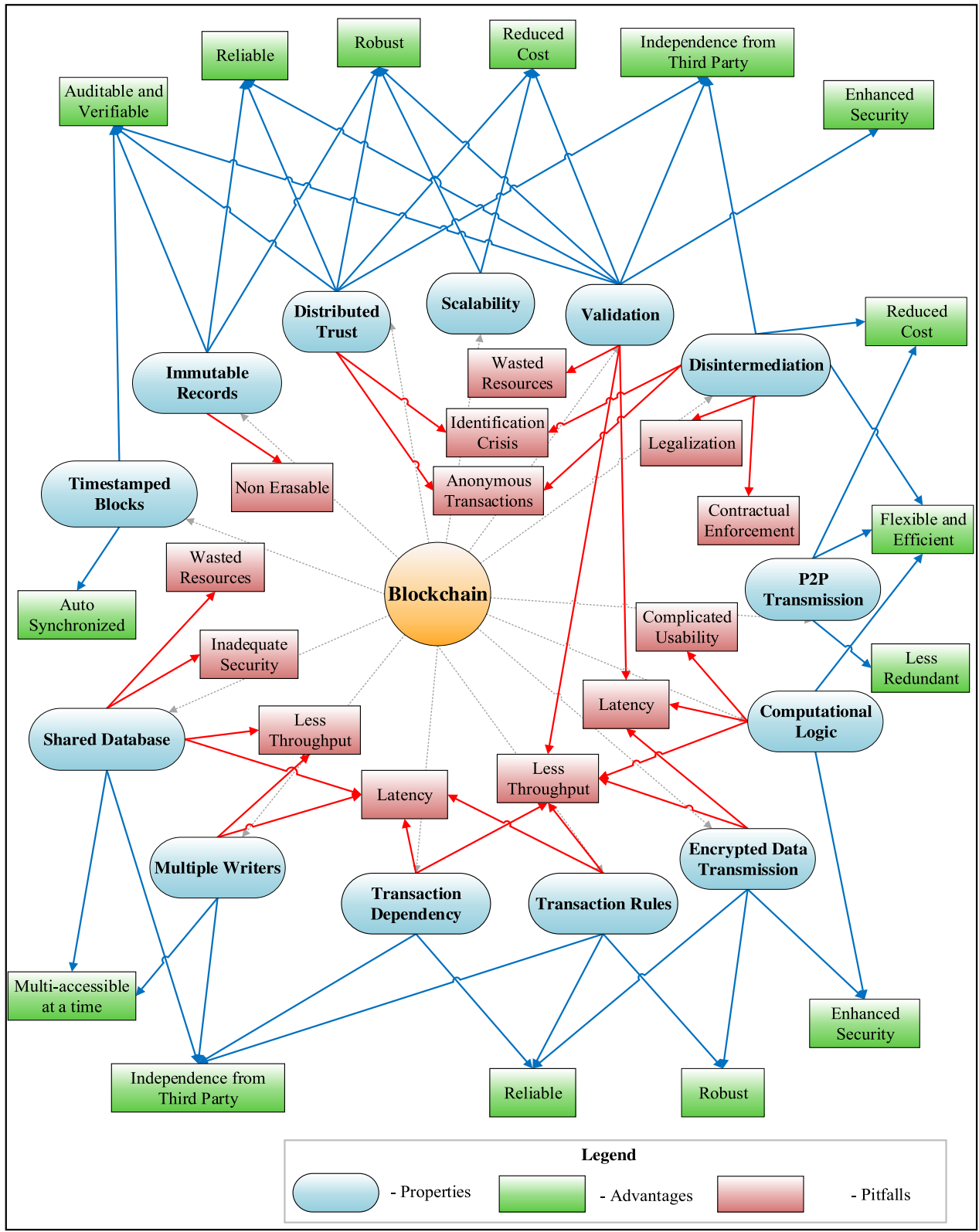

**FIGURE 3.** Mapping among the properties, advantages and pitfalls of blockchain.

Service perspective refers to the useful properties offered by blockchain, which includes *scalability, validation, multiple writers* and *distributed trust*. Logical inclusion is the combination of the logical properties such as *computational logic, transaction dependency* and *transaction rules*. Architectural characteristics refer to the structural features, which

**TABLE 6.** Association among the properties, advantages and pitfalls of blockchain.

| Properties | | Independence from third party | Enhanced Security | Flexible & efficient | Auditable & verifiable | Robust | Reduced cost | Auto Synchronized | Less redundant | Reliable | Multi-accessible |
|---|---|---|---|---|---|---|---|---|---|---|---|
| | | **Advantages** | | | | | | | | | |
| | Shared Database | ✓ | | ✓(r) | ✓(r) | ✓ | ✓(r) | | | | ✓ |
| | P2P Transmission | | | ✓ | | | ✓ | | ✓ | | |
| | Timestamped blocks | | | | ✓ | | | ✓ | | | |
| | Immutable Records | | | | ✓ | ✓ | | | ✓(r) | ✓ | |
| | Encrypted data transmission | | ✓ | | ✓(r) | ✓ ✓(r) | | | | ✓ | |
| | Disintermediation | ✓ | ✓(r) | ✓ | | | ✓ ✓(r) | | | ✓(r) | ✓(r) |
| | Computational Logic | ✓(r) | ✓ | ✓ | ✓(r) | ✓ | | | | | |
| | Transaction Dependency | ✓ | ✓ | | ✓(r) | ✓ | | | | ✓ | |
| | Transaction Rules | ✓ | ✓ | | ✓(r) | ✓ ✓(r) | | | | ✓ | |
| | Distributed Trust | | | | ✓ | ✓ | | | | ✓ ✓(r) | ✓(r) |
| | Multiple Writers | ✓ | | | ✓(r) | ✓ | | | | ✓(r) | ✓ ✓(r) |
| | Validation | ✓ | ✓ | ✓(r) | ✓ ✓(r) | ✓ ✓(r) | ✓ | | | | |
| | Scalability | | | | | ✓ | ✓ | | | | |
| | | Complicated Usability | Legalization | Wasted Resources | Latency | Less Throughput | Contractual Enforcement | Inadequate Security | Non Erasable | Anonymous Transaction | Identification Crisis |
| | | **Pitfalls** | | | | | | | | | |

*(Green tick ✓ = advantage achieved by the property; red tick ✓(r) = pitfall.)*

include *shared database, P2P transmission, disintermediation, timestamped blocks, immutable records* and *encrypted data transmission*. The properties are presented in a propositional table (see Table 3) and a conceptual map derived from the table is depicted in Fig. 2.

### C. ADVANTAGES AND PITFALLS OF BLOCKCHAIN

A total of ten advantages and ten pitfalls were found through the literature survey and presented in Table 4 and Table 5, respectively. The advantages and pitfalls were presented with respect to the context (i.e. the area where the article focused on). For example, *independence from third party* was found as an advantage in four articles in the context of smart city, bitcoin and medical. *Enhanced security* (in case of finance, smart city, and software license validation), *flexible and efficient* (in economics), and *auditable and verifiable* (in the context of economics, medical and software license validation) were mentioned in three articles as benefits.

In case of pitfalls, *complicated usability* (in case of e-governance), *legalization* (in the context of e-governance and economics), *wasted resources* (in economics) and *latency* (in economics) are mostly mentioned pitfalls (three times).

We also note that many of the advantages and pitfalls are seen as trade-offs. For example, in order to improve throughput, reduce electricity consumption and increase sustainability, one needs to use a mining algorithm other than the proof-of-work. This, in turn, reduces security. Therefore, parameters to be considered in implementation of a blockchain for an application are highly dependent on the context. A benefit in one context may turn as a pitfall in another context. In the case of bitcoin, it seems that the community prefers security over sustainability and throughput [2]. However, due to the high consumption of electricity, this approach may not be viable in the long run. Furthermore, due to the lower throughput, bitcoin would probably fail to gain wider adoption.

## V. MAPPING AMONG PROPERTIES, ADVANTAGES, AND PITFALLS OF BLOCKCHAIN

A mapping among the properties, advantages, and pitfalls of blockchain has been presented in Fig. 3. Each node of the concept map represents a concept. The blockchain is at the center. The blue oval shapes are properties, the green and red rectangles are advantages and pitfalls, respectively. The properties are linked to benefits. Links from multiple properties to a benefit means all those properties may provide that particular advantage. The figure also shows the association of properties, benefits, and pitfalls.

For example, *independence from third party* is an advantage, which can be achieved from four properties of blockchain: *distributed database*, *distributed control*, *decentralization* and *mined blocks* [3], [7], [11], [57]. On the other hand, *anonymous transactions* is a pitfall, which can be provided by three properties including *encrypted data transmission*, *transparency with pseudonymity* and *decentralization* [15], [22].

The association or mapping among the properties, advantages, and pitfalls has been depicted in Table 6. Here, the green tick mark (✓) represents that the advantage can be achieved by the properties and similarly, the red tick mark (✓) is for the pitfalls.

A few benefits were also found as pitfalls depending on how blockchain is being utilized. (See Fig. 3 and Table 6). For example, *enhanced security* can be considered as an advantage since all the transactions use public key cryptography for data transmission [7], [11], [15]; whereas *inadequate security* can be considered as a pitfall since there is high possibility of cyber-attack in blockchain based systems [1], [17].

## VI. CONCLUSION

Blockchain is a relatively newer innovation and most of the research studies have mainly focused on proposing conceptual frameworks and algorithms. In the industrial field, the implementation of blockchain is still very limited except in the area of cryptocurrencies.

In this research, a rigorous literature survey has been conducted to define the blockchain technology in a more systematic way including its possible features and expected benefits and pitfalls. As outcome, a few concept maps and a mapping among the properties, benefits, and pitfalls of blockchain have been provided.

The outcome of this research will provide fundamental knowledge facilitating the future researchers to integrate blockchain in their development of future technological solutions. The mapping among the properties, advantages, and pitfalls provides a clear understanding for better utilization of blockchain.

We note two limitations of our research. First, as some specific strings are used to search related literature, a part of related articles may be missed. Second, the mapping of the features with benefits and pitfalls was based on the existing research. The properties that have not been studied in any field are not included in our concept maps.

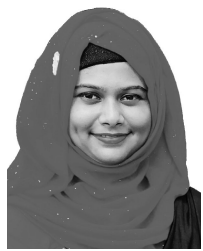
**IYOLITA ISLAM** received the B.Sc. degree in computer science and engineering from the Military Institute of Science and Technology (MIST), Dhaka, Bangladesh, in 2017, where she is currently pursuing the M.Sc. degree in computer science and engineering. She is serving as a Lecturer with the Computer Science and Engineering Department, MIST, since 2018. She is the author of several conference papers. Her research interests include human–computer interaction, blockchain technology, health informatics, and industry 4.0. She was awarded the Best Paper Award in International Conference on Sustainable Technologies for Industry 4.0 (STI 2019). She is an Associate Member of The Institution of Engineers, Bangladesh (IEB).

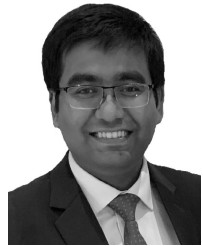
**KAZI MD. MUNIM** received the B.Sc degree in computer science and engineering from the Military Institute of Science and Technology (MIST), Dhaka, Bangladesh, in 2017, where he is currently pursuing the M.Sc degree. He is currently a Software Quality Assurance Engineer by Profession. His research interests include human–computer interaction, blockchain, and industry 4.0. He is the author of several conference papers. He was awarded the Best Paper Award in International Conference on Sustainable Technologies for Industry 4.0 (STI 2019).

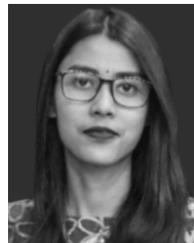
**SHAHRIMA JANNAT OISHWEE** received the B.Sc. degree in computer science and engineering from the Military Institute of Science and Technology (MIST), Dhaka, Bangladesh, in 2017. Since January 2018, she has been serving as a Lecturer with the Computer Science and Engineering Department, MIST. She is the author of several conference papers. Her research interests include human–computer interaction, data science in the context of information security, software security, quality, and usability analysis. She is currently an Associate Member of The Institution of Engineers, Bangladesh (IEB).

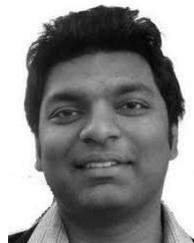
**A. K. M. NAJMUL ISLAM** received the M.Sc. (Eng.) degree from Tampere University of Technology, Finland, and the Ph.D. degree in information systems from the University of Turku, Finland. He is currently an Adjunct Professor with Tampere University, Finland. He also works as University Research Fellow with the Department of Future Technologies, University of Turku. He has 80+ publications. His research focuses on Human Centered Computing. His research has been published in top outlets such as Information Systems Journal, *Journal of Strategic Information Systems, Technological Forecasting and Social Change, Computers in Human Behavior, Internet Research, Computers and Education, Information Technology and People, Telematics and Informatics, Communications of the AIS, Journal of Information Systems Education, AIS Transaction on Human-Computer Interaction*, and *Behaviour and Information Technology*, among others.

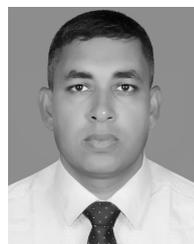
**MUHAMMAD NAZRUL ISLAM** received the B.Sc. degree in computer science and information technology from the Islamic University of Technology, Bangladesh, in 2002, the M.Sc. degree in computer engineering from Politecnico di Milano, Italy, in 2007, and the Ph.D. degree in information systems from Åbo Akademi University, Finland, in 2014. He is currently an Associate Professor with the Department of Computer Science and Engineering, Military Institute of Science and Technology (MIST), Mirpur Cantonment, Dhaka, Bangladesh. Before joining MIST, he was working as a Visiting Teaching Fellow with Uppsala University, Sweden and as a Postdoctoral Research Fellow with Åbo Akademi University, Finland. He was also a Lecturer and an Assistant Professor with the Department of Computer Science and Engineering, Khulna University of Engineering and Technology (KUET), Bangladesh, from 2003 to 2012. His research areas include but not limited to human–computer interaction (HCI), humanitarian technology, health informatics, military information systems, information systems usability, and computer semiotics. He is the author of more than 80 peer-reviewed publications in International journals and conferences. He is a member of The Institution of Engineers, Bangladesh (IEB).

. . .